\documentclass{ws-procs9x6}

\usepackage[absolute]{textpos}

\usepackage{xspace, units} 

\newcommand {\bfI} {{\bf I}\xspace}
\newcommand {\bfA} {{\bf A}\xspace}

\newcommand {\cs} {$\mathcal{C}^S$\xspace}
\newcommand {\cc} {$\mathcal{C}^C$\xspace}
\newcommand {\cb} {$\mathcal{C}^B$\xspace}

\begin{document}

\title{ON THE CENTRALITY OF THE FOCUS IN HUMAN EPILEPTIC BRAIN NETWORKS}

\author{G. GEIER\textsuperscript{1,2,*}, M.-T. KUHNERT\textsuperscript{1,2,3}, C. E. ELGER\textsuperscript{1}, K. LEHNERTZ\textsuperscript{1,2,3}}

\address{
    \textsuperscript{1}Department of Epileptology, University of Bonn,\\ Sigmund-Freud-Stra{\ss}e~25, 53105~Bonn, Germany\\
    \textsuperscript{2}Helmholtz-Institute for Radiation and Nuclear Physics, University of Bonn,\\ Nussallee~14--16, 53115~Bonn, Germany\\
    \textsuperscript{3}Interdisciplinary Center for Complex Systems, University of Bonn,\\ Br{\"u}hler Stra{\ss}e~7, 53175~Bonn, Germany\\
    \textsuperscript{*}E-mail: \texttt{geier@uni-bonn.de}
        }

\begin{abstract}
There is increasing evidence for specific cortical and subcortical large-scale human epileptic networks to be involved in the generation, spread, and termination of not only primary generalized but also focal onset seizures. 
The complex dynamics of such networks has been studied with methods of analysis from graph theory. In addition to investigating network-specific characteristics, recent studies aim to determine the functional role of single nodes---such as the epileptic focus---in epileptic brain networks and their relationship to ictogenesis. 
Utilizing the concept of betweenness centrality to assess the importance of network nodes, previous studies reported the epileptic focus to be of highest importance prior to seizures, which would support the notion of a network hub that facilitates seizure activity. 
We performed a time-resolved analysis of various aspects of node importance in epileptic brain networks derived from long-term, multi-channel, intracranial electroencephalographic recordings from an epilepsy patient.
Our preliminary findings indicate that the epileptic focus is not consistently the most important network node, but node importance may drastically vary over time.
\end{abstract}

\begin{textblock*}{14cm}(3cm,27cm)
    \noindent R. Tetzlaff and C. E. Elger and K. Lehnertz (2013), \emph{Recent Advances in Predicting and Preventing Epileptic Seizures}, page 175--185, Singapore, World Scientific.\\
    Copyright 2013 by World Scientific.
\end{textblock*}

\bodymatter

\section{Introduction}
Over the last decade network analysis has proven to be an invaluable tool to advance our understanding of complex dynamical systems in diverse scientific fields\cite{Strogatz2001, Albert2002, Newman2003, Boccaletti2006a, Arenas2008, Fortunato2010, Newman2012} including the neuroscienes\cite{Reijneveld2007, Bullmore2009, Sporns2011a, Stam2012}. 
Specific aspects of functional brain networks---with nodes that are usually associated with sensors capturing the dynamics of different brain regions and with links representing interactions\cite{Pikovsky2001, Kantz2003, Pereda2005, Hlavackova2007, Marwan2007, Lehnertz2009b, Friedrich2011, Lehnertz2011b} between pairs of brain regions---were reported to differ between epilepsy patients and healthy controls\cite{Chavez2010, Horstmann2010, Ansmann2012} which supports the concept of an epileptic network\cite{Bertram1998, Bragin2000, Spencer2002, Lemieux2011, Berg2011}.
Moreover, epileptic networks during generalized and focal seizures (including status epilepticus) were shown to possess topologies that differ from those during the seizure-free interval\cite{Ponten2007, Schindler2008a, Ponten2009, vanDellen2009, Kramer2010, Kuhnert2010, Bialonski2011b, Kramer2011, Gupta2011}. 
Most of the aforementioned studies investigated network-specific characteristics such as the average shortest path length or the clustering coefficient. 
Network theory, however, also provides concepts and tools to assess various aspects of importance (e.g. centralities) of a node in a network\cite{Freeman1979, Bonacich1987, Koschutzki2005, Estrada2010, Kuhnert2012}, but by now, there are only a few studies that investigated node-specific characteristics of epileptic networks\cite{Kramer2008, Wilke2011, Varotto2012}, and these studies investigated the dynamics of functional brain networks during seizures only. Refs. \refcite{Wilke2011} and \refcite{Varotto2012} reported on highest centrality values for the (clinically defined) epileptic focus which would support the notion of a 
crucial network node
that facilitates seizure activity.

We here report preliminary findings obtained from a time-resolved analysis of node importance in functional brain networks derived from long-term, multi-channel, intracranial electroencephalographic (iEEG) recordings from an epilepsy patient.
Investigating various centrality aspects, we provide first evidence that the epileptic focus is not consistently the most important node (i.e., with highest centrality), but node importance may drastically vary over time.

\section{Methods}
\subsection{Inferring Weighted Functional Networks}
We analyzed iEEG data from a patient who underwent presurgical evaluation of drug-resistant epilepsy of left mesial-temporal origin
and who is completely seizure free after selective amygdalohippocampectomy.
The patient had signed informed consent that the clinical data might be used and published for research purposes. 
The study protocol had previously been approved by the local ethics committee. 
iEEG was recorded from $N=60$ channels (chronically implanted intrahippocampal depth and subdural grid and strip electrodes) and the total recording time amounted to about 
1.7 days, during which three seizures were observed.
iEEG data were sampled at \unit[200]{Hz} using a \unit[16]{bit} analog-to-digital converter, filtered within a frequency band of \unit[0.1--70]{Hz}, and referenced against the average of two
recording contacts outside the focal region. 

Following previous studies\cite{Horstmann2010, Kuhnert2010, Kuhnert2012, Ansmann2012} we associated each recording site with a network node and defined functional network links between any pair of nodes $j$ and $k$---regardless of their anatomical connectivity---using the mean phase coherence $R_{j,k}$ as a measure for signal interdependencies\cite{Mormann2000}.
We used a sliding window approach with non-overlapping windows of $M=4096$ data points (duration: \unit[20.48]{s}) each to estimate $R_{j,k}$ in a time-resolved fashion, employing the Hilbert transform to extract the phases $\Phi$ from the windowed iEEG. The elements of the interdependence matrix \bfI then read:

\begin{equation}
	\label{eq:R}
R_{jk}=
\left|\left(\frac{1}{M}\sum_{m=0}^{M-1}{\exp i\left(\Phi_j(m)-\Phi_k(m)\right)}\right)\right|.
\end{equation}

In order to derive an adjacency matrix \bfA from \bfI (i.e, an undirected, weighted functional network) and to account for the case that the centrality metrics could reflect trivial properties of the weight collection\cite{Ansmann2011} we sort $\left\{R_{jk} \;\middle|\;  j < k \right\}$ in ascending order and denote with $\upsilon_{jk}$ the position of $R_{jk}$ in this order (rank). 
We then consider $A_{jk}=2\upsilon_{jk}/(N(N-1))$, $j \neq k$, and $A_{jj}=0$. 
This approach leads to a weight collection with entries being uniformly distributed in the interval $[0,1]$.

\subsection{Estimating Centrality}

The importance of a network node may be assessed via centrality metrics\cite{Freeman1979, Bonacich1987, Koschutzki2005, Estrada2010}.
Degree, closeness, and betweenness centrality are frequently used for network analyses, and for these metrics generalizations to weighted networks have been proposed (see Ref. \refcite{Kuhnert2012} for an overview).

If a node is adjacent to many other nodes, it possesses a high degree centrality. 
When investigating weighted networks, however, the number of neighboring nodes is not a sensible measure and one may consider {\em strength centrality} of node $j$ instead\cite{Barrat2004b} 

\begin{equation}
\mathcal{C}^S(j) = \frac{\sum_{k}{a_{jk}}}{N-1}.
\label{eq:cs}
\end{equation}

Assessing node importance in weighted networks via closeness and betweenness centrality requires the definition of shortest paths. This can be achieved by assuming the ``length'' of a link to vary inversely with its weight\cite{Newman2004}. The {\em closeness centrality} of node $j$ is defined as

\begin{equation}
\mathcal{C}^C(j) = \frac{N-1}{\sum_k{d_{jk}}},
\label{eq:cc}
\end{equation}
where $d_{jk}$ denotes the length of the shortest path from node $j$ to node $k$. 

The {\em betweenness centrality} of node $j$ is the fraction of shortest paths running through that node.
\begin{equation}
   \mathcal{C}^B(j) = \frac{2}{(N-1)(N-2)} \sum_{h=0}^{N}\sum_{\substack{
   k=0\\ k\neq j}}^{N}\frac{\eta_{hk}(j)}{\eta_{hk}}.
 \label{eq:bc}
 \end{equation}

\begin{figure}[ht] \centering
    \includegraphics[width=0.7\columnwidth]{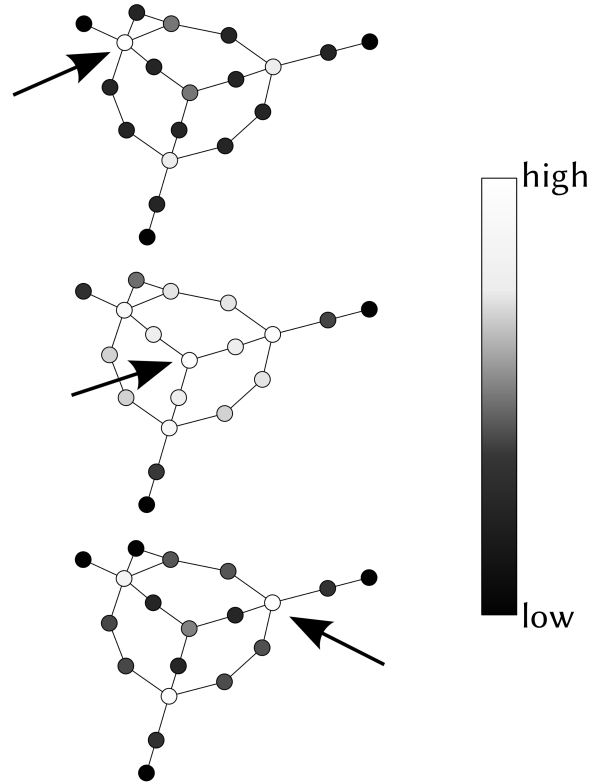}
\caption{
Values of degree centrality (top), closeness centrality (middle) and betweenness centrality (bottom) for nodes of an exemplary binary network. The most important node (highest centrality) is indicated by an arrow.}
\label{img:centr}
\end{figure}

Here, $\eta_{hk}(j)$ denotes the number of shortest paths between nodes $h$ and $k$ running through node $j$, and $\eta_{hk}$ is the total number of shortest paths between nodes $h$ and $k$. 
We used the algorithm proposed by Brandes\cite{Brandes2001} to estimate the aforementioned centralities. Fig. \ref{img:centr} illustrates the centrality metrics  \cs, \cc, and \cb for the nodes of an exemplary network.

\section{Results}

\begin{figure}[ht] \centering
\includegraphics[width=0.9 \columnwidth]{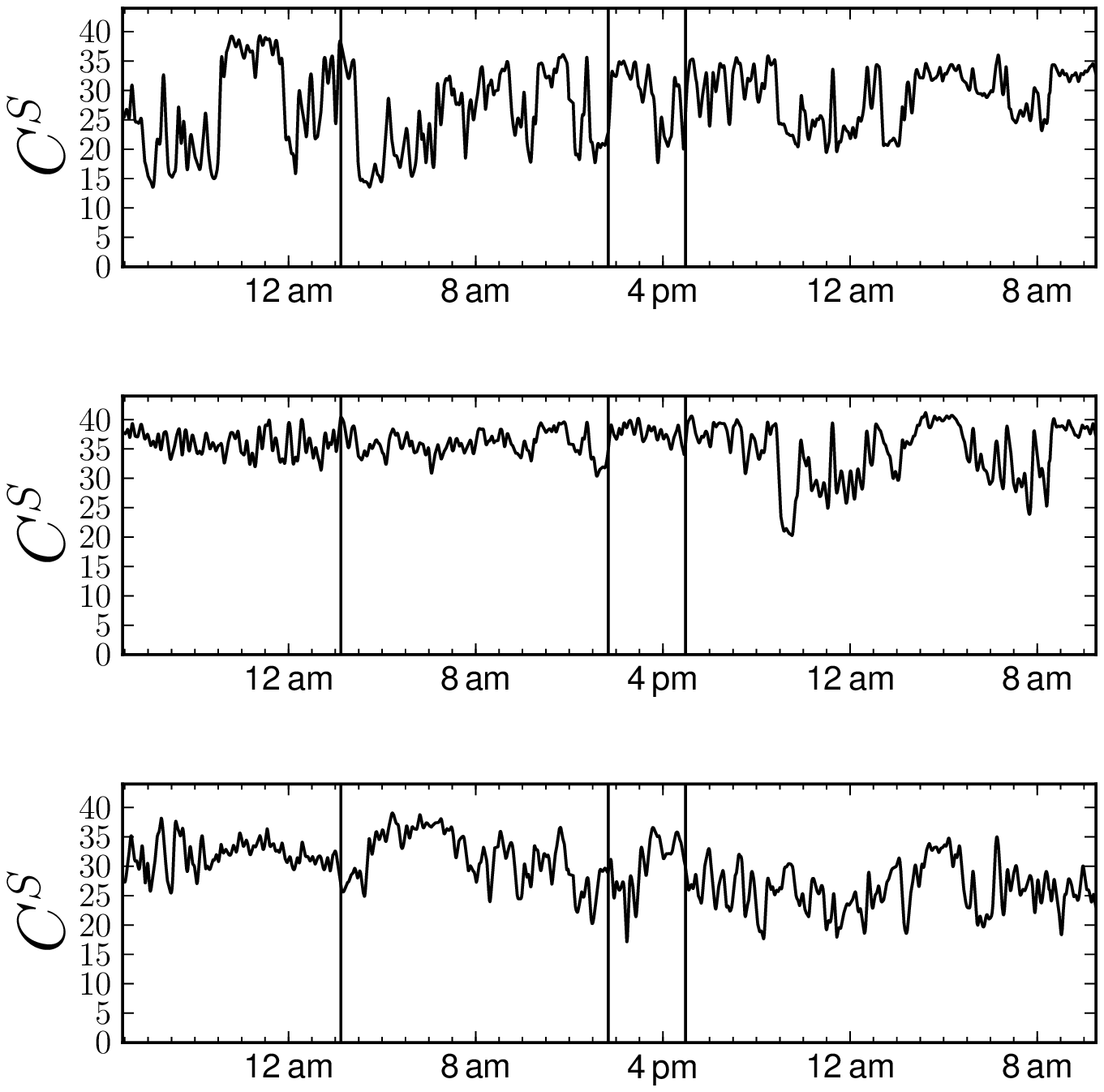}
\caption{Temporal evolution of strength centrality for a node located within the clinically defined epileptic focus (top), for a node located in its immediate surrounding (middle), and for a distant node (bottom). Recording time was \unit[41]{h}, during which three seizures occurred. Moving average over 4096 windows corresponding to \unit[20.48]{s}. Black vertical lines mark the times of electrical seizure onsets. For legibility, all curves were smoothed using a Gaussian kernel ($\sigma = \unit[5]{min}$).}
\label{img:dc_example}
\end{figure}

\begin{figure}[ht] \centering
\includegraphics[width=0.9 \columnwidth]{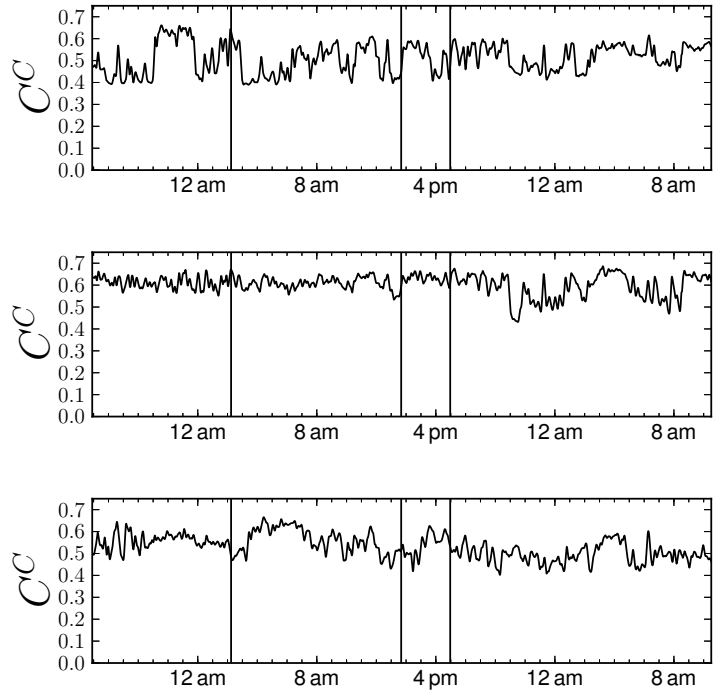}
\caption{Same as Fig. \ref{img:dc_example} but for closeness centrality.}
\label{img:cc_example}
\end{figure}

\begin{figure}[ht] \centering
\includegraphics[width=0.9 \columnwidth]{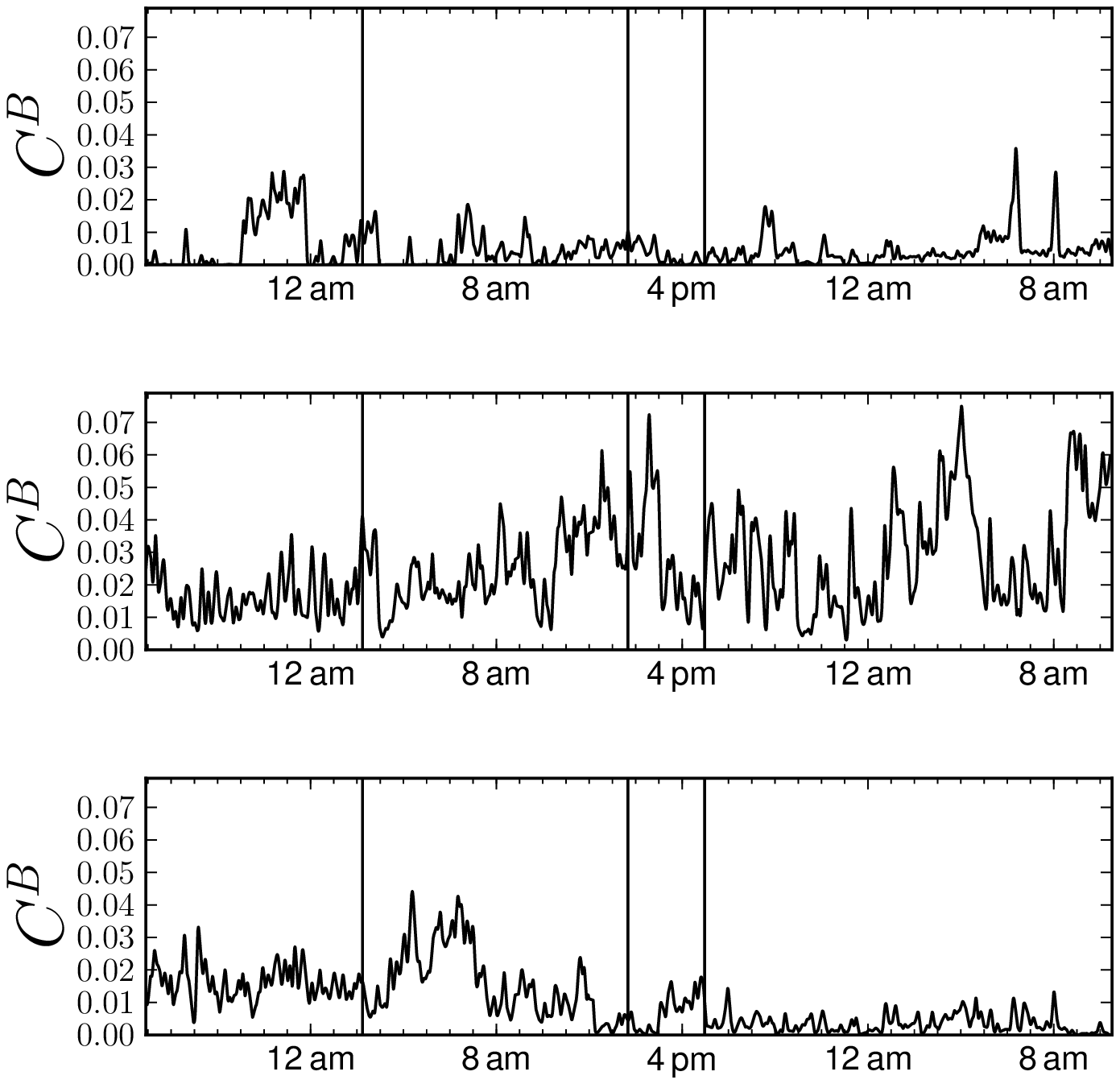}
\caption{Same as Fig. \ref{img:dc_example} but for betweenness centrality.}
\label{img:bc_example}
\end{figure}

In Figs. \ref{img:dc_example}, \ref{img:cc_example}, and \ref{img:bc_example} we show the temporal 
evolutions of \cs, \cc and \cb over \unit[41]{h} for three selected nodes from the exemplary epileptic brain networks investigated here. 
We chose one node from within the epileptic focus (upper plots of figures), another node from the immediate surrounding of the epileptic focus (middle plots of figures), and a third one which was associated with a recording site far off the epileptic focus (lower plots in figures). 
All centrality metrics exhibited large fluctuations over time, both on shorter and longer time scales. 
The temporal evolutions of \cs and \cc were quite similar, while \cb behaved differently from the two other metrics. 
The similarity between \cs and \cc was to be expected, at least to some degree (see the discussion in Ref. \refcite{Kuhnert2012}), since they characterize the role of a node as a starting or end point of a path. On the other hand, \cb characterizes a node's share of all paths between pairs of nodes that utilize that node. 

For this patient, we could not observe any clear cut changes of the centrality metrics prior to seizures that would indicate a preictal state. 
Moreover, none of the metrics exhibited features in their temporal evolutions that would constantly indicate the network nodes associated with the epileptic focus (or its immediate neighborhood) as important nodes. Rather, their importance may drastically vary over time.

\begin{figure}[ht] \centering
\includegraphics[width=1 \columnwidth]{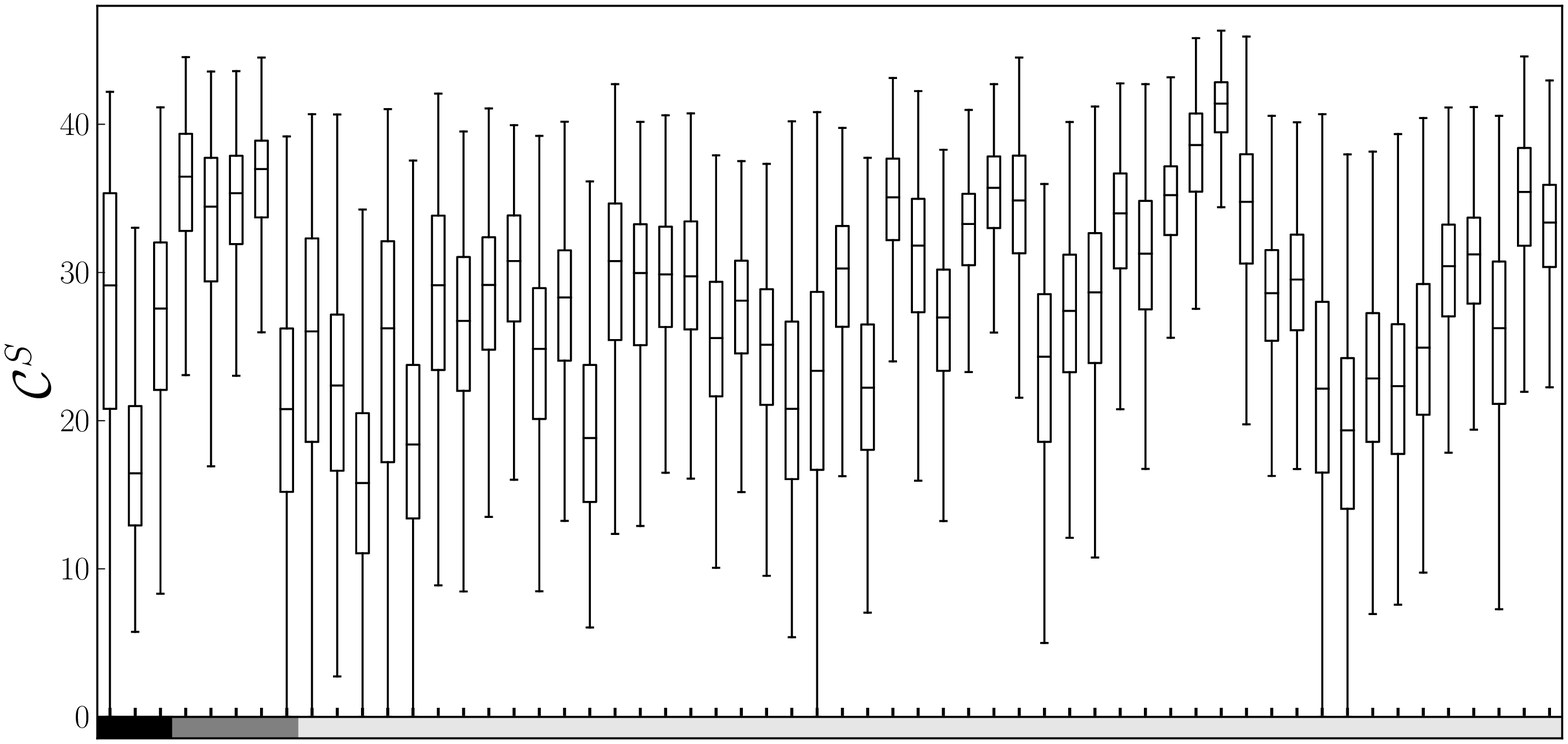}
\includegraphics[width=1 \columnwidth]{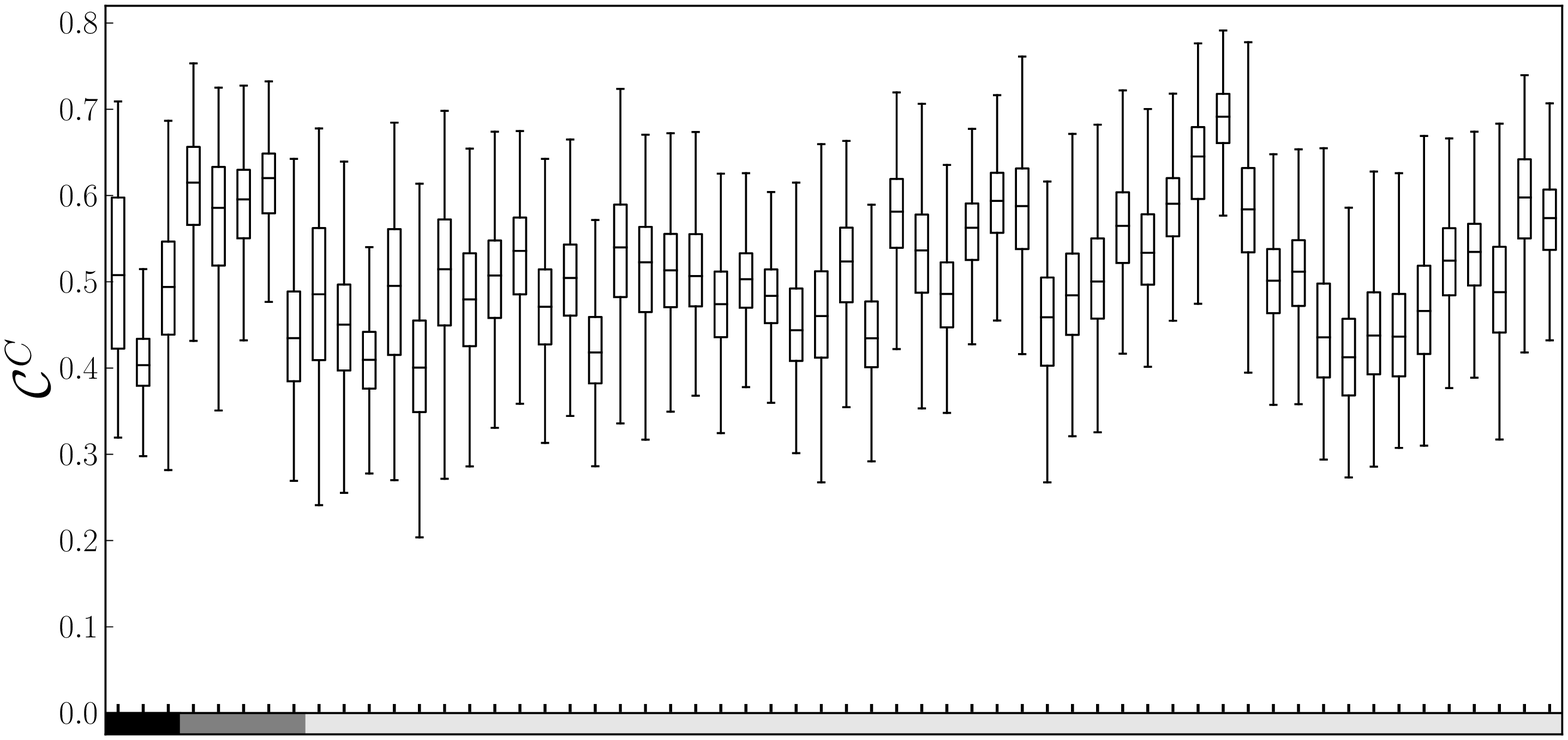}
\includegraphics[width=1 \columnwidth]{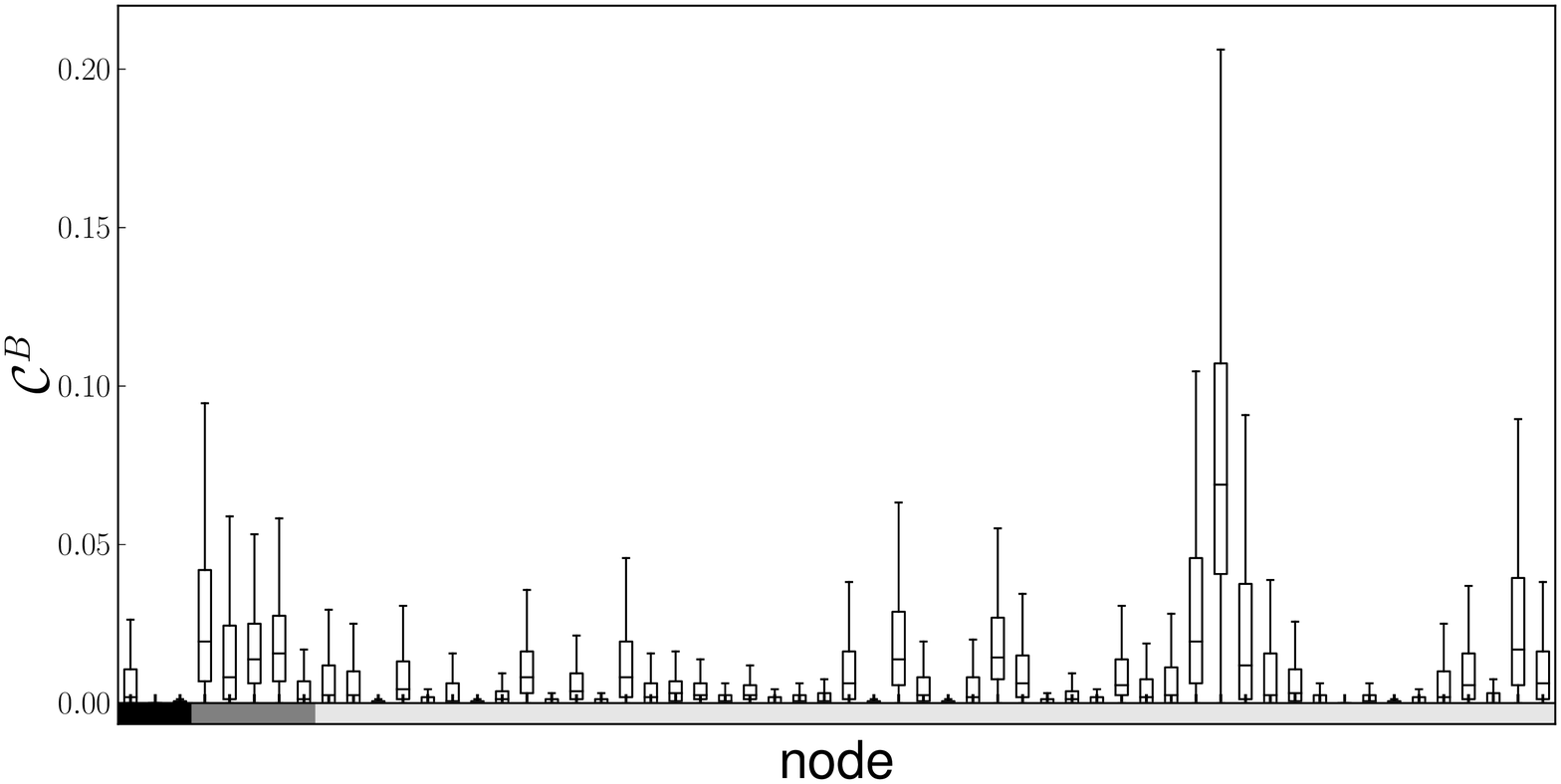}
\caption{Statistical characteristics of \cs (top) , \cc (middle), and \cb (bottom) of each node depicted as boxplot.
Dashed error bars extend from the sample minimum to the sample maximum.
Bottom and top of a box are the lower and upper quartile, and the vertical line in the box denotes the median.
The black bar on the abscissa indicates nodes associated with the clinically defined epileptic focus, the grey bar indicates nodes associated with the immediate surrounding of the epileptic focus, and the white bar indicates distant nodes.
}
\label{img:boxplots}
\end{figure}

To demonstrate that our exemplary results hold for all nodes of the epileptic brain networks investigated here, we show, in Fig. \ref{img:boxplots}, findings obtained from an exploratory data analysis. The main statistical characteristics of centralities of each node (maximum and minimum value, the median, and the quartiles estimated from the respective temporal evolutions) indicated that neither the epileptic focus nor its immediate surrounding can be considered as important, and that the different centrality metrics ranked different nodes as most important.

\section{Conclusion}
We have investigated various aspects of centrality of individual nodes in epileptic brain networks derived from long-term, multi-channel iEEG recordings from an epilepsy patient. 
Utilizing different centrality metrics, we observed nodes far from the clinically defined epileptic focus and its immediate surrounding to be the most important ones. 
Although our findings must, at present, be regarded as preliminary, they are nevertheless in stark contrast to previous studies\cite{Wilke2011, Varotto2012} that reported highest node centralities for the epileptic focus only. 
It remains to be investigated whether the different findings can be attributed to the dynamics of
different epileptic brains or to, e.g., differences in network inference. 
One also needs to take into account that there are a number of potentially confounding variables whose impact on estimates of different centrality metrics is still poorly understood. 

\section*{Acknowledgments}
This work was supported by the Deutsche Forschungsgemeinschaft (Grant No. LE660/4-2).

\bibliographystyle{ws-procs9x6}

\end{document}